\documentclass[letter]{jpsj2} 
\title{Impurity Effect as a Probe for the Gap Function in  the Filled Skutterudite Compound Superconductor PrOs$_{4}$Sb$_{12}$:
Sb-NQR Study }

\author{
Masahide \textsc{Nishiyama}$^{}$,
Takayuki \textsc{Kato}$^{}$,
Hitoshi \textsc{Sugawara}$^{1}$,
Daisuke \textsc{Kikuchi}$^{2}$,
Hideyuki \textsc{Sato}$^{2}$,
Hisatomo \textsc{Harima}$^{3}$,
and Guo-qing \textsc{Zheng}$^{}$
}

\inst{
$^{}$Department of Physics, Okayama University,
Okayama 700-8530, Japan \\
$^{1}$Department of Mathematical and Natural Sciences,
Faculty of Integrated Arts and Sciences,
The University of Tokushima, Tokushima 770-8502, Japan \\
$^{2}$Department of Physics, Graduate School of Science, Tokyo Metropolitan University, Hachioji 192-0397, Japan \\
$^{3}$Department of Physics,  Kobe University, Kobe 657-8501, Japan }
\abst{
We have carried out nuclear quadrupole resonance (NQR) measurements in the filled skutterudite compounds Pr(Os$_{1-x}$Ru$_x$)$_4$Sb$_{12}$ ($x=0.1, 0.2$), in order to gain insights into the symmetry of the superconducting gap function. Upon replacing Os with Ru,
the spin-lattice relaxation rate $1/T_1$ becomes proportional to temperature ($T$) at low $T$ far below $T_c$, and the magnitude of $(1/T_1T)_{low-T}$ increases with increasing Ru content. These results indicate that a finite density of states is induced at the Fermi level by the impurity, and thus suggest that there exist  nodes in the gap function of PrOs$_{4}$Sb$_{12}$.
}

\kword{filled skutterudite, Heavy fermion, superconductivity,  impurity effect, nuclear quadrupole resonance, residual DOS}

\begin{document}
\maketitle


The superconductivity at $T_c$=1.85 K  discovered in the filled skutterudite compound PrOs$_{4}$Sb$_{12}$ has attracted much attention \cite{firstPr_HFSC1,firstPr_HFSC2}. This is the 
first Pr-based heavy fermion superconductor, in which the heavy mass has been suggested by the large specific
heat jump $\Delta C/T_{\rm c}$$\sim$500~mJ/K$^{2}\cdot$ mol at  $T_{\rm
c}$,~\cite{firstPr_HFSC1,firstPr_HFSC2} and directly confirmed by  de Haas-van
Alphen (dHvA) experiments \cite{Sugawara}. The crystal electric field (CEF) ground state is a $\Gamma_{1}$ singlet, which is separated by the first excited state of the $\Gamma_{4}^{(2)}$ triplet by a gap of  $\Delta_{CEF}$=10 K \cite{Aoki,Tenya,Tayama,Kohgi,Kuwahara,Goremychkin}. Because of this small $\Delta_{CEF}$,  the role of quadrupole moment fluctuations arising from the $\Gamma_{4}^{(2)}$ state has attracted much attention. It has been speculated that the interaction between the quadrupolar moments  and the 
 conduction-electron charges may be responsible for the heavy mass, thus representing a charge-scattering version of the Kondo effect \cite{Cox}. The contrasting behavior that the isostructural compound  PrRu$_{4}$Sb$_{12}$ with $\Delta_{CEF}$=70 K has a much smaller $m^*$  \cite{Takeda,Matsuda} has increased such expectation.
 
 One of the first steps toward understanding the Pr-based heavy fermions  may be the determination of the gap symmetry of the superconductivity.
 Previous nuclear quadrupole resonance (NQR) measurements have revealed the unconventional nature of the superconductivity \cite{PrOsSbNQR}. No coherence peak was found in the spin-lattice relaxation rate $1/T_1$ just below $T_c$, which suggests non-BCS-type superconductivity and is in contrast to the $s$-wave bahavior subsequently observed in PrRu$_4$Sb$_{12}$ \cite{Yogi}. At low temperatures, $1/T_1$ does not follow a power-law $T$-dependence, which is in contrast to the observation of a $T^3$ variation in other known heavy fermion superconductors \cite{Kitaoka}. Together with the exponential decrease in the penetration depth found by muon spin relaxation \cite{Mac}, it has been suggested that the superconducting gap is isotropic.  
 However, oscillations with respect to the magnetic field angle have been found in angle-resolved thermal conductivity measurements, which suggests that the gap is anisotropic \cite{PrOsSbIzawa}.
 
 In this study, we use an impurity as a probe for the gap function. We replace Os with Ru, and study its effect on the superconducting density of states (DOS) using the NQR technique. It has been known for some time that a non-magnetic impurity is a powerful "smoking gun" for the gap function. In $s$-wave superconductors,  a paramagnetic impurity produces a bound state within the gap (Yu-Shiba state), \cite{Yu,Shiba} and non-magnetic impurities smear out the gap anisotropy, if any. \cite{Norman} In contrast, if there are nodes in the gap function, a finite DOS is brought about at the Fermi level \cite{Miyake1,Miyake2}, which can be detected experimentally \cite{Kitaoka}.

Single crystals of Pr(Os$_{1-x}$Ru$_{x}$)$_{4}$Sb$_{12}$, ($x$=0.1 and 0.2) were grown by the Sb-flux method. $T_c$'s are 1.6 K and 1.4 K for $x$=0.1 and 0.2, respectively, in agreement with those previously reported \cite{PrOsSbFrederick1,Chia}.
For NQR measurements, 
the single crystals were  powdered in order to allow the radio-frequency magnetic field to penetrate into the sample. The NQR measurements were performed by using a phase coherent spectrometer. 
The spin-lattice relaxation rate, $1/T_{1}$, for $^{123}$Sb were measured using the saturation-recovery method. Data below 1.4 K were collected using a $^3$He/$^4$He dilution refrigerator.

\begin{figure}[tb]
\begin{center}
\includegraphics[width=9cm]{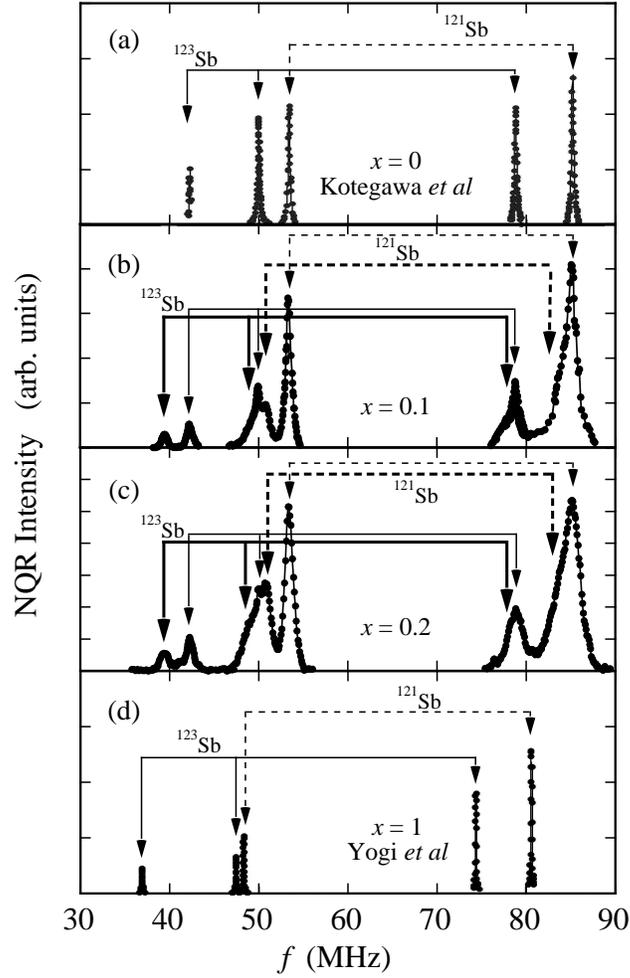}
\end{center}
\caption{Sb-NQR spectra of Pr(Os$_{1-x}$Ru$_{x}$)$_{4}$Sb$_{12}$ at 4.2 K. 
The data for $x=0$ and  $x=1$ are from Kotegawa {\it et al.} \cite{PrOsSbNQR} and Yogi {\it et al} \cite{Yogi}.  In (b) and (c), the spectra indicated by the thin solid lines (broken lines) correspond to $^{123}$Sb ($^{121}$Sb) signals seen in PrOs$_{4}$Sb$_{12}$, while the spectra indicated by the thick markers are the new sets arising from Ru doping.
}
\label{f1}
\end{figure}

Figure 1 shows the NQR spectra for the $x$=0.1 and 0.2 samples. For comparison, the data for the two end-member compounds  \cite{PrOsSbNQR,Yogi} are also shown in Figs 1(a) and 1(d). In the Ru-doped samples, in addition to the spectrum peaks that are located at the same positions as those for pure PrOs$_4$Sb$_{12}$, there appear new sets of peaks that are characterized by $\nu_Q$=26 MHz and the asymmetry parameter $\eta$=0.43 for $^{123}$Sb. Here, $\nu_{Q}$ and $\eta$ are defined as $\nu_{Q}\equiv\nu_{z}=\frac{3}{2I(2I-1)h}e^2Q\frac{\partial ^2V}{\partial z^2}$ and $\eta=|\nu_{x}-\nu_{y}|/\nu_{z}$  with   $Q$ being the nuclear quadrupolar moment and $\frac{\partial ^2V}{\partial \alpha^2} (\alpha=x, y, z)$ being the electric field gradient  at the position of the nucleus \cite{Abragam}.
It should be emphasized that these peaks (labelled as ``peaks 2" in Table I) are different from those for pure PrRu$_4$Sb$_{12}$. These new peaks are due to the alloying effect, which indicates that the alloying is homogenous and there is no phase separation. Such a feature has been seen before in other heavy fermion compounds. For example, in Ir-doped CeRhIn$_{5}$, upon replacing Rh with Ir, there emerges a new set of NQR peaks that is different from that for CeRhIn$_5$ or CeIrIn$_5$ \cite{Zheng}. Nonetheless, $1/T_1$ measured at the new set of peaks shows the same $T$-dependence as that measured at those corresponding to CeRhIn$_5$ or CeIrIn$_5$, indicating that  the new set of peaks is not due to phase-separated regions but rather to the homogenous alloying. 

\begin{table}[tb]
\caption{NQR frequency $\nu_{Q}$ and asymmetric 
parameter $\eta$ for $^{123}$Sb in 
Pr(Os$_{1-x}$Ru$_{x}$)$_{4}$Sb$_{12}$. Also shown are the CEF gap $\Delta_{{\rm CEF}}$ and the heavy quasiparticle enhancement factor $\beta$ (see text).}
\label{eqQ}
\setlength{\doublerulesep}{0.8pt}
\begin{tabular}{|c||cc|cc||cc|}
\hline
& \multicolumn{2}{c|}{Peaks 1} & \multicolumn{2}{c||}{Peaks 2} & 
\multicolumn{2}{c|}{\,} \\
\cline{2-5}
$x$ &$\nu_{Q}$\,(MHz) &$ \eta$ & $\nu_{Q}$\,(MHz)&$ \eta$&$\beta$& 
$\Delta_{{\rm CEF}}$ (K)\\
\hline
0   &26.8&0.41&---&---&26.7&8\\
0.1& 26.8& 0.41& 26& 0.43&16&10\\
0.2& 26.8& 0.41& 26& 0.43&11&13\\
1    &25.2&0.46&---&---&---&---\\
\hline
\end{tabular}
\end{table}%


The nuclear spin-lattice relaxation rate $1/T_1$ was measured at the peaks at 49.9\,MHz ($\pm3/2\leftarrow \rightarrow \pm5/2$ transition) and 42.2 MHz ($\pm1/2\leftarrow \rightarrow \pm3/2$ transition).  Both measurements yield the same results. Even for $x$=0.2, no influence on $T_1$ due to the overlapping of the $^{121}$Sb line was seen in the measurements at the 49.9\,MHz peak, probably because the $H_1$ we used for observing the echo is much smaller than the frequency  difference between the $^{121}$Sb and $^{123}$Sb peaks, which is about 0.8 MHz, and the $^{121}$Sb peak has a small full width at half maximum (FWHM) of 0.5 MHz. Above $T\sim$ 100 K, the nuclear magnetization can  be fitted excellently using the expected theoretical rate equation \cite{Chepin}. However, below $T\sim$ 100 K, it cannot be fitted using the theoretical curve with a single component of $T_1$.   This is also true  at the peak at 42.2 MHz that is free from overlapping by other transition lines. Figure 2 shows the decay curve of the nuclear magnetization at $T$=6 K. As will become   clearer later, the inhomogeneous $T_1$ is an intrinsic property of the alloyed samples. The one-component behavior for $T\geq$ 100 K is rather a coincidence, since $T_1$  is the same for both PrOs$_4$Sb$_{12}$ and PrRu$_4$Sb$_{12}$.

\begin{figure}[tb]
\begin{center}
\includegraphics[width=7cm]{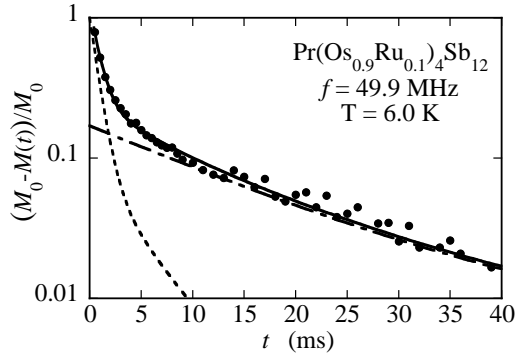}
\end{center}
\caption{
Nuclear magnetization recovery curve. The broken and  dot-dashed curves indicate the recovery curves due to the fast and slow relaxation components, respectively.
The solid curve indicates the sum of the two components.
}
\label{f2}
\end{figure}

We then attempt to fit the low-$T$ nuclear magnetization with two $T_1$ components, namely, 

\begin{eqnarray}
\frac{M_{0}-M(t)}{M_{0}} =&\nonumber \\
& \sum_{i=S,L} a_i\left[0.0762  \exp \left( -\frac{3t}{T_{1}^{i}}\right)
+0.0146  \exp \left( -\frac{8.9t}{T_{1}^{i}}\right)
+0.9092  \exp \left( -\frac{17.3t}{T_{1}^{i}}\right) \right], 
\label{}
\end{eqnarray}
where $T_1^S$ means the short component due to fast relaxation, and $T_1^L$, the long component due to slow relaxation. 
The solid curve in Fig. 2 shows such fitting for $x$=0.1, with $a_S$=0.8 and $a_L$=0.2  that do  not depend on temperature appreciably. 

\begin{figure}[tb]
\begin{center}
\includegraphics[width=8cm]{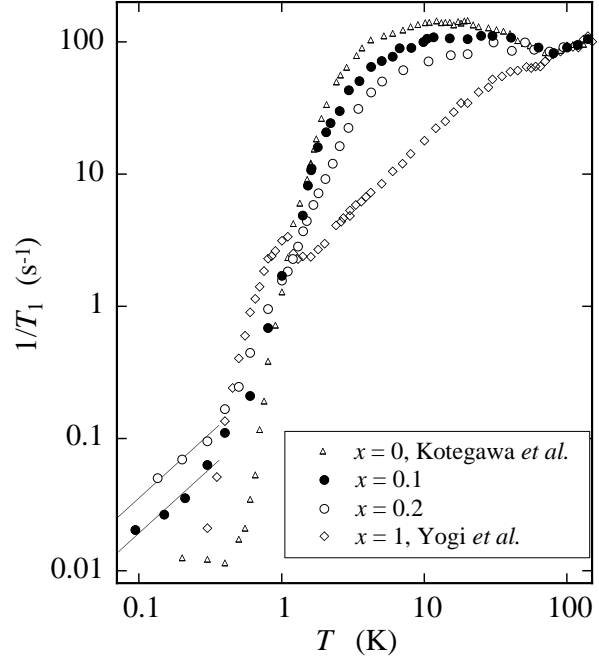}
\end{center}
\caption{Temperature dependence of $1/T_1^S$ in Pr(Os$_{1-x}$Ru$_x$)$_4$Sb$_{12}$ ($x=0.1, 0.2$). Data for   PrOs$_4$Sb$_{12}$ \cite{PrOsSbNQR} and  PrRu$_4$Sb$_{12}$, which is a BCS superconductor \cite{Yogi}, are also shown for comparison. The straight lines indicate the $T_1T$=const relations.}
\label{f3}
\end{figure}

We first discuss the main component, $1/T_1^S$, whose temperature dependence   is shown in Fig. 3. Compared to the results for the end-member compounds, several trends can be seen. First, above $T\sim$ 100 K, there is only one component of $T_1$ (namely, $T_1^S=T_1^L$), and the data for all $x$ merge into a same line. This suggests that  the relaxation at high $T$ is {\it{not}} governed by the electronic state that is sensitive to the transition metal element, \cite{Harima} but due to other degree of freedom. Second, below $T\sim$ 100 K, $1/T_1^S$ is close to that for pure PrOs$_4$Sb$_{12}$, but decreases as $x$ increases.  Since the slow component $1/T_1^L$ is very close to that for pure PrRu$_4$Sb$_{12}$, as  will be shown later,  we consider that $1/T_1^S$ origenates from the Sb sites that are located far from Ru. Third, and most importantly, below $T\sim$ 1 K, $1/T_1^S$ is proportional to $T$; the magnitude of $1/T_1^ST$ increases with increasing $x$. 

The  $1/T_1^ST$=const. relation indicates that a finite DOS is induced by the impurity. $1/T_1$ in the superconducting state may be expressed as 
\begin{eqnarray}
\frac{T_1(T=T_c)}{T_{1}} = \frac{2}{k_BT_c}\int (N_s(E)^{2}+M_s(E)^{2})f(E)(1-f(E))dE,
\end{eqnarray}
where $N_{s}(E)=N_{0} E/(E^2-\Delta^2)^{1/2}$ is the superconducting DOS with $\Delta$ being the superconducting gap,  $M_{s}(E)=N_{0} \Delta/(E^2-\Delta^2)^{1/2}$ is the anomalous DOS due to the coherence factor \cite{MacL}, $N_{0}$ is the DOS in the normal state and $f(E)$ is the Fermi function. In the $T$=0 limit, $1/T_1$ is dominated by $N_s(0)^2$. The present results indicate that $N_s(0)$ is finite in the Ru-doped samples and increases with increasing $x$. This result cannot be accounted for by an $s$-wave gap \cite{Yu,Shiba,Norman}.
It is emphasized that only if there is a sign change in the gap function, namely, there are nodes in the gap function, can the impurity induce the finite DOS \cite{Norman}. Thus, our results provide strong evidence for the existence of nodes in the gap function. The seemingly exponential decrease in  $1/T_1$ with respect to $T$ in pure PrOs$_4$Sb$_{12}$ may be an effect  obscured by the double superconducting transitions \cite{Vollmer,Measson}.
Let us now estimate the impurity-induced  $N_s(0)$, denoted as $N_{res}$ hereafter. The values of $N_{res}$ calculated from the relation of
\begin{eqnarray}
\frac{N_{res}}{N_0} = \sqrt{\frac{(T_1^sT)_{T_c}}{(T_1^sT)_{low-T}} }
\end{eqnarray}
are shown in Fig. 4. The curves indicate the calculated results obtained by Miyake  for the gaps with line-nodes (axial) and point-nodes (polar) \cite{Miyake1,Miyake2}. The experimental results agree qualitatively  with the theoretical results. That the experimental data fall below the theoretical curve is probably due to the depression of the pairing force by the substitution of Ru, which is not included in the theoretical results. We will discuss this point later. Although we are unable to distinguish between line nodes and point nodes in Fig. 4, point nodes nonetheless seem more plausible,
in view of the relatively weaker suppression of $T_c$ by the impurity with respect to $x$ \cite{PrOsSbFrederick1}.  
\begin{figure}[tb]
\begin{center}
\includegraphics[width=8cm]{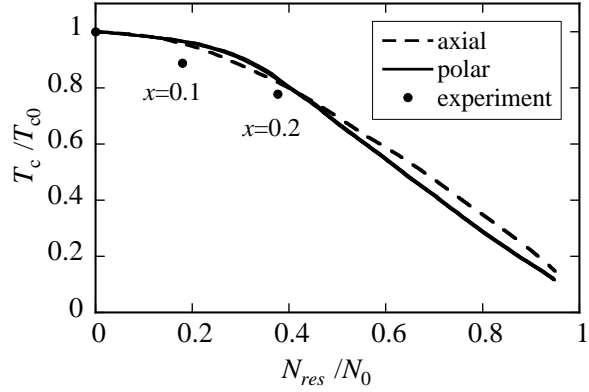}
\end{center}
\caption{Suppression of $T_c$ as a function of the residual DOS induced by the impurity, where $T_{c0}$ is the transition temperature in the absence of impurity scattering. The curves indicate the  calculated results obtained by Miyake\cite{Miyake1,Miyake2}. }
\label{f4}
\end{figure}

For completeness, we show the temperature dependence of $1/T_1^L$ in Fig. 5. Above $T_c$, $1/T_1^L$ shows a temperature variation very close to that of PrRu$_4$Sb$_{12}$. This $1/T_1^L$ can be assigned to come from the Sb sites that are located close to Ru. Below $T_c$, $1/T_1^L$ decreases rapidly, 
 with no coherence peak. 

\begin{figure}[tb]
\begin{center}
\includegraphics[width=7cm]{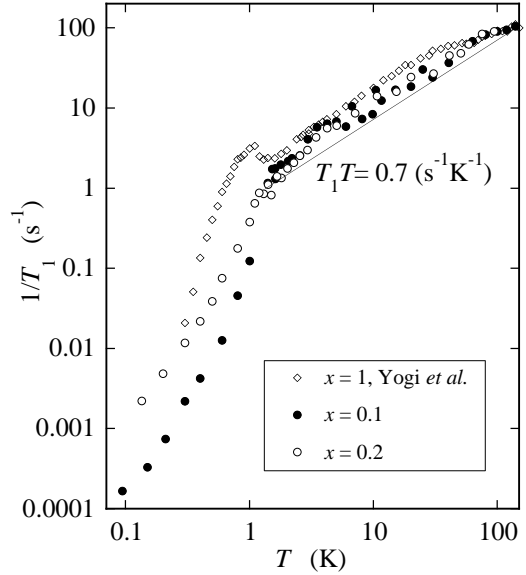}
\end{center}
\caption{Temperature dependence of the minor, slow relaxation component, $1/T_1^L$. The straight line indicates the $1/T_1T$=0.7 s$^{-1}$K$^{-1}$ relation  found in LaOs$_4$Sb$_{12}$.}
\label{f5}
\end{figure}

Finally, we discuss the high-temperature behavior of $1/T_1^S$. Figure 6 shows the temperature variation of the quantity $1/T_1T$. Below $T\sim$ 6 K, $1/T_1T$ decreases, leaving a peak at around $T\sim$ 6 K. Such a decrease becomes mild as $x$ increases.
Since the CEF gap $\Delta_{CEF}$ is small, Kotegawa {\it et al.} have analyzed the low-$T$ data by decomposing $1/T_1T$ into two parts \cite{PrOsSbNQR}, namely, the contribution due to the excitation to the  $\Gamma_{4}^{(2)}$ state, and that due to heavy  quasiparticles: 
$1/T_1T=\alpha \times exp(-\Delta_{CEF}/k_BT) +\beta \times 0.7$ s$^{-1}$K$^{-1}$, 
where 0.7 s$^{-1}$K$^{-1}$ is the $1/T_1T$ for LaOs$_4$Sb$_{12}$ and the parameter $\sqrt{\beta}$ represents the mass enhancement factor of the  heavy quasiparticles.  The application of the same analysis  of Kotegawa {\it et al.} to the $x$=0.1 and 0.2 samples, as shown in the inset of Fig. 6, shows that $\Delta_{CEF}$ increases with increasing $x$, while $\beta$ decreases with increasing $x$. The results are summarized in Table I, which are in good agreement with those reported by  Frederick {\it et al.} who deduced $\Delta_{CEF}$ from the susceptibility and the mass enhancement factor from the specific heat coefficient $\gamma$ \cite{PrOsSbFrederick1}. As has been discussed  by several authors \cite{Yogi,PrOsSbFrederick1}, the increase in $\Delta_{CEF}$ is responsible, at least partly, for the decrease in $T_c$ in going from PrOs$_4$Sb$_{12}$ to PrRu$_4$Sb$_{12}$; thus, the discrepancy between the theoretical curve and the experimental data seen in Fig. 4 can be ascribed to the increase in $\Delta_{CEF}$.

\begin{figure}[tb]
\begin{center}
\includegraphics[width=7cm]{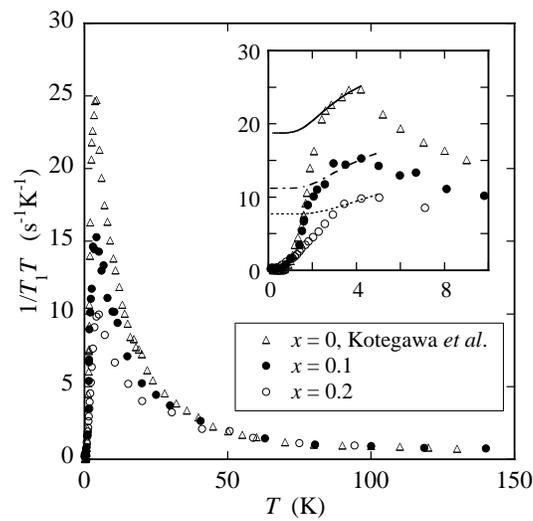}
\end{center}
\caption{Temperature dependence of the quantity $1/T_1^ST$. The inset shows the enlarged part below $T$=10 K. The curves indicate the results of fitting $1/T_1^ST=\alpha \times exp(-\Delta_{CEF}/k_BT) +\beta \times 0.7$ s$^{-1}$K$^{-1}$. For details, see the text.}
\label{f6}
\end{figure}



In conclusion, we have studied the impurity effect on the superconductivity in the filled skutterudite heavy fermion superconductor PrOs$_4$Sb$_{12}$ using the NQR technique. We find that replacing Os with Ru brings about a finite density of states at the Fermi level, which increases with increasing Ru content. Our results provide strong evidence for the existence of nodes in the gap function.






\section*{Acknowledgment}
We are grateful to Y. Kitaoka, H. Kotegawa, Yogi and Y. Imamura  for helpful discussions and contribution. We  would also like to thank  K. Miyake for providing the unpublished calculated results shown in Fig. 4.  This work was supported in part by a research grant from MEXT on the Priority Area ``Skutterudites" (No. 15072204). 



\end{document}